\documentstyle[twoside,fleqn,espcrc2]{article}

\newcommand{\Pdy}[1]{\frac{\partial}{\partial #1}}
\newcommand{\func}[2]{#1\!\left(\!#2\!\right)}
\newcommand{\txtfrac}[2]{\phantom{1\!\!\!}^{#1}\!\!/\!_{#2}}

\begin{document}

\title{Lattice HQET Calculation of the Isgur-Wise Function}

\author{Joseph Christensen\thanks{Presented by J.\ Christensen at
         Lattice '97, Edinburgh, Scotland.},
	Terrence Draper
    and Craig McNeile\thanks{Currently at Department of Physics, 
	 University of Utah, Salt Lake City, UT 84112.}
        \address{Department of Physics and Astronomy, University of 
	 Kentucky, Lexington, KY 40506}
	\thanks{This work is supported in part by the U.S. Department
         of Energy under grant numbers DE-FG05-84ER40154 and
         DE-FC02-91ER75661, and by the University of Kentucky Center
         for Computational Sciences.  The computations were carried
         out at NERSC.}}

\begin{abstract}
We calculate the Isgur-Wise function on the lattice, simulating the
light quark with the Wilson action and the heavy quark with a direct
lattice implementation of the heavy-quark effective theory.  Improved
smearing functions produced by a variational technique, {\sc most},
are used to reduce the statistical errors and to minimize
excited-state contamination of the ground-state signal.  Calculating
the required matching factors, we obtain $\xi'(1)=-0.64(13)$ for the
slope of the Isgur-Wise function in continuum-HQET in the
$\overline{\rm MS}$ scheme at a scale of $4.0\,$GeV.
\end{abstract}

\maketitle


\section{The Tadpole-Improved Simulation}
\label{s:simulation}

The Isgur-Wise function is the form factor of a heavy-light meson in
which the heavy quark is taken to be much heavier than the energy
scale, $m_Q^{} \gg \Lambda_{\rm QCD}^{}$.  This calculation adds
perturbative corrections to the simulation results of Draper \&
McNeile~\cite{Draper95a0}.  The Isgur-Wise function is calculated
using the action first suggested by Mandula \&
Ogilvie~\cite{Mandula92}:
\begin{eqnarray}
iS 
& \!\!=\!\!
    & \sum_x \left\{ v_0 \left[ \func{\psi^{\dagger}}{x} \func{\psi}{x}
    - \func{\psi^{\dagger}}{x} \frac{\func{U_t}{x}}{u_0} 
      \func{\psi}{x+\hat{t}}
      \right]
      \right. \nonumber \\ & & \left. \phantom{\sum}
    + \sum_{j=1}^3 \frac{-i v_j }{ 2 } 
      \left[ \func{\psi^{\dagger}}{x} \frac{\func{U_j}{x}}{u_0} 
             \func{\psi}{x+\hat{j}} 
	     \right. \right. \nonumber \\ & & \left. \left. \phantom{\sum}
           - \func{\psi^{\dagger}}{x} 
             \frac{U_j^{\dagger}\!(x-\hat{j})}{u_0} 
             \func{\psi}{x-\hat{j}} \right] 
      \right\}
      \nonumber 
\end{eqnarray}
This leads to the evolution equation:
\begin{eqnarray}
G(x\!+\!\hat{t}) = 
     \frac{U^{\dagger}_t(x)}{u_0} 
      \left\{ G(x) \phantom{\frac{\func{U_j}{x}}{u_0}}
              \ \ \ \ \ \ \ \ \ \ \ \ 
              \ \ \ \ \ \ \ \ \ \ \ \ 
              \right. \!\! \!\! \!\! \!\! \!\! \!\! & & \nonumber \\ \left. 
            - \sum_{j=1}^3 \frac{i \tilde{v}_j}{ 2 } 
                           \left[ \frac{\func{U_j}{x}}{u_0} 
                                  G(x\!+\!\hat{j})
                                - \frac{U_j^\dagger\!(x\!-\!\hat{j})\!}
                                       {u_0}
                                  G(x\!-\!\hat{j}) \right] \right\}
      \!\! \!\! \!\! \!\! \!\! \!\! 
&   & \nonumber  
\end{eqnarray}
where $\tilde{v}_j = \frac{v_j}{v_0}$ and $G(\vec{x},t=0) =
\frac{1}{v_0} f(\vec{x})$.  Better results can be obtained when a
smeared source is used in place of a point source.  The smearing
function, $f(\vec{x})$, was calculated from a static simulation
($v_j=0$) {\it via\/} the smearing technique {\sc most} (Maximal
Operator Smearing Technique~\cite{Draper94a0}).


\section{The Isgur-Wise Function}
\label{s:isgur-wise}

The Isgur-Wise function was extracted from the lattice simulation as
a ratio of three-point functions which was suggested by Mandula \&
Ogilvie~\cite{Mandula92}.  $\left| \xi^{\rm lat}_{\rm unren}(v \cdot
v') \right|^2$ is the large $\Delta t$ limit of
\begin{eqnarray}
\!\!\!\! & \!\!\!\!\!\! & \!\!
\frac{4 v_0 v'_0}{(v_0\!+\!v'_0)^2}
      \frac{ C_3^{vv'}(\Delta t) C_3^{v'v}(\Delta t) }
           { C_3^{vv}(\Delta t) C_3^{v'v'}(\Delta t) }
      \nonumber
\end{eqnarray}
where $\Delta t$ is the time separation between the current operator
and each $B$-meson interpolating field.

Draper \& McNeile have presented~\cite{Draper95a0} the
non-tadpole-improved unrenormalized slope of the lattice Isgur-Wise
function to demonstrate the efficacy of the computational techniques.


\section{Tadpole Improvement for HQET}
\label{s:tadpole}

Tadpole improvement grew from the observation that lattice links,
$U$, have mean field value, $u_0 \ne 1$.  Therefore, it is better to
use an action written as a function of ($\txtfrac{U}{u_0}$).  In the
Wilson action, each link has a coefficient $\kappa$; $u_0$ can be
paired with $\kappa$ to easily tadpole improve {\it a posteriori\/}
any previous non-tadpole-improved calculation.

In the HQET, there is no common coefficient (analogous to $\kappa$)
for both $U_t$ and $U_j$.  Correspondence between tadpole-improved
and non-tadpole-improved HQET actions with $v^2=1$ cannot be made
{\it via\/} a simple rescaling of parameters, as is done for the
Wilson action with $\kappa$.

Fortunately, the evolution equation \underline{can\/} be written (as
noticed by Mandula \& Ogilvie~\cite{Mandula96/97}) such that the
$u_0$ is grouped with $\tilde{v}_j$.  Thus, tadpole-improved
Monte-Carlo data can be {\it constructed\/} from the
non-tadpole-improved data by replacing $v^{\rm nt} \rightarrow v^{\rm
tad}$ and by including two overall multiplicative factors:
\begin{eqnarray}
G^{\rm tad}(t;\tilde{v}^{\rm tad}, v_0^{\rm tad})
& \!\!=\!\! & u_0^{-t} \frac{v_0^{\rm nt}}{v_0^{\rm tad}}
                       G^{\rm nt}(t;\tilde{v}^{\rm nt}, v_0^{\rm nt})
              \label{eq:two-point}
\end{eqnarray}


In addition to the multiplicative factors $u_0^{-t}$ and
$\txtfrac{v_0^{\rm nt}}{v_0^{\rm tad}}$, the tadpole-improvement of
the simulation requires adjusting the velocity according to
$\tilde{v}^{\rm tad} = u_0 \tilde{v}^{\rm nt}$, subject to $(v^{\rm
tad})^2 = 1$ and $(v^{\rm nt})^2 = 1$.  Thus,
\begin{eqnarray}
v^{\rm tad}_0 
& = & v^{\rm nt}_0 [1+(1-u_0^2) (v^{\rm nt}_j)^2]^{-\txtfrac{1}{2}}
      \nonumber \\
v^{\rm tad}_j 
& = & u_0^{} v^{\rm nt}_j [1+(1-u_0^2) (v^{\rm nt}_j)^2]^{-\txtfrac{1}{2}}
      \nonumber
\end{eqnarray}
%


\section{Tadpole-Improved Renormalization}
\label{renorm1}

By comparing the unrenormalized propagator
\begin{eqnarray}
\left[ v_0^b \left( \frac{e^{i k_4}}{u_0}  - 1 \right)
     + \frac{v_z^b}{u_0} \sin(k_z) + M_0^b - \Sigma(k,v) \right]^{-1}
\!\! \!\! 
\!\! \!\! \!\! 
& &  \nonumber
\end{eqnarray}
to the renormalized propagator
\begin{eqnarray}
i H(k,v) & = & Z_Q \left[ v_0^r (ik_4) + v_z^r ( k_z) + M^r \right]^{-1}
               \nonumber
\end{eqnarray}
at $O(k^2)$ and using $(v^r)^2 \!=\! (v^b)^2 \!=\! 1$, the
perturbative renormalizations can be obtained.
Aglietti~\cite{Aglietti94} has done this for a different
non-tadpole-improved action, for the special case $\vec{v} = v_z
\hat{z}$.

With momentum shift, $p \!\!\rightarrow\!\! p' \!\!=\!\!  \left<
p_4\!+\!i\ln(\!u_0\!), \vec{p} \right>$ \cite{Bernard94a0}, with
$\txtfrac{1}{u_0} \exp(i k_4) = \exp(i(k_4+i\ln(u_0)))$ and with
$X_\mu \equiv \Pdy{p_\mu} \left. \Sigma(p) \right|_{p=0}$, the
tadpole-improved perturbative renormalizations are found to be
\begin{eqnarray}
\delta M 
& = & -\Sigma(0) - v_0 \ln(u_0) 
      \nonumber \\
\delta Z_Q
& = & Z_Q - 1 \ = \ 
     -iv_0 X_4 
    - u_0 \sum_{j=1}^{3} v_j X_j 
      \nonumber \\
\delta^{\underline{v_i}}_{u_0}
& = & -iv_0 \frac{v_i}{u_0} X_4 
    - (1+v_i^2) X_i - v_i \sum_{j \ne i} v_j X_j 
      \nonumber \\
\delta v_0
& = & -i\sum_{j=1}^{3} v_j^2 X_4 
    - u_0 v_0 \sum_{j=1}^3 v_j X_j 
      \nonumber
\end{eqnarray}
$u_0$ is the perturbative expansion and~\footnote{Note: $(v^2=1)
\Rightarrow \left( v_0 \delta v_0 = u_0^2 \sum_j
\phantom{1\!\!}^{\underline{v_j}}_{u_0}
\delta^{\underline{v_j}}_{u_0}\right)$.}
\begin{eqnarray}
v^{r,\rm tad}_j
  =   v_j^{b,\rm tad}Z_{v_j}^{\rm tad} 
& , & 
v^{r,\rm tad}_0
  =   v^{b,\rm tad}_0 Z_{v_0}^{\rm tad} 
      \nonumber \\
Z_{v_j}^{\rm tad} 
\equiv \frac{1}{u_0} 
       \left( 1 
            + \frac{\delta^{\underline{v_j}}_{u_0}}
                   {\phantom{1\!\!}^{\underline{v_j}}_{u_0}} \right)
& , &
Z_{v_0}^{\rm tad} 
\equiv 1 + \frac{\delta v_0}{v_0} 
       \nonumber
\end{eqnarray}
If one fits to $\exp\{-t\}$ rather than $\exp\{-(t+1)\}$, the
tadpole-improved wave-function renormalization is reduced to $\delta
Z'_Q = \delta Z_Q + (\delta M^{\rm tad} + v_0 \ln(u_0))/v_0$.  Thus
the $\ln(u_0)$ term cancels explicitly and, as in the static
case~\cite{Bernard94a0}, tadpole-improvement has no effect on $\delta
Z'_Q$, to order $\alpha$.


%
\section{Perturbative Renormalizations}

We will present our computations of the renormalization factors
elsewhere, but include this comment: Although the tadpole-improved
functions include factors of $\left. u_0 \right|_{\rm
pt}$~\cite{Lepage93}, these effects are higher order in $\alpha$ and
are dropped.  Only the velocity renormalization is explicitly
affected by the perturbative expansion of $u_0$:
\begin{eqnarray}
Z_{v_j}^{\rm tad} 
& = & \left( 1 
           + \frac{\delta^{\underline{v_j}}_{u_0}}
                  {\phantom{1\!\!}^{\underline{v_j}}_{u_0}}
           - \frac{g^2 C_F}{16 \pi^2} (-\pi^2)
             \right)
      \nonumber 
\end{eqnarray}

The perturbative renormalizations favor a scale of $q^*a = 1.9(1)$
for $\alpha$, which yields $\alpha \approx 0.19(1)$.


\section{Velocity Renormalization}

Mandula \& Ogilvie~\cite{Mandula96/97} consider the perturbative
velocity renormalization expanded in orders of $\tilde{v}$.  Our
numbers for the velocity renormalization agree with theirs.

Another option is to consider, as did both Mandula \& Ogilvie and
Hashimoto \& Matsufuru~\cite{Hashimoto96}, the non-perturbative
renormalization of the velocity.  From Hashimoto's \& Matsufuru's
graph, we estimate their $Z_{v,\rm np}^{\rm tad} \approx 1.05(5)$.
From Mandula \& Ogilvie's result, we notice that $Z_{\tilde{v},\rm
np}^{\rm nt} = u_0 \times 1.01(1)$.  This is very close to the effect
of tadpole-improving, and implies $Z_{\tilde{v}}^{\rm tad} =
1.01(1)$.

We therefore use $Z_v^{\rm tad} \approx 1$ as the
non-per\-tur\-ba\-tive velocity renormalization in our calculation to
renormalize the slope of the Isgur-Wise function.


\section{Renormalization of $\xi'(v \cdot v')$}

We claim that we can convert our Monte-Carlo data into
tadpole-improved results and can calculate a renormalized
tadpole-improved slope for the Isgur-Wise function.

We use the notation $Z'_\xi = 1 + \delta Z'_\xi$ for the
renormalization of the Isgur-Wise function, with $\delta Z'_\xi$:
\begin{eqnarray}
\frac{g^2}{12\pi^2} 
\left[ 2 \left( 1 - v \!\cdot\! v' r(v \!\cdot\! v') \right)
       \ln (\mu a)^2 
     - f'(v,v') \right] \nonumber
\end{eqnarray}
with $r(v \!\cdot\! v')$ defined in~\cite{falk90} and primes on $Z$
and $f$ to indicate the ``reduced value.''

For simplicity, we use the local current, which
is not conserved on the lattice; $Z'_\xi(1) \ne 1$.  However, the
construction in \S\ref{s:isgur-wise} guarantees that the extracted
renormalized Isgur-Wise function is properly normalized, $\xi_{\rm
ren}(1)=1$.


\section{Conclusions}
\begin{table}[t]
\begin{center}
\begin{tabular}{cllll} 
\multicolumn{5}{c}{\bf Unrenormalized Isgur-Wise Slope} \\ \hline \hline
\multicolumn{5}{c}{\bf Not Tadpole Improved} \\
\hline
$\Delta t$ & 0.154 & 0.155 & 0.156 
           & \multicolumn{1}{c}{$\kappa_c$} \\
\hline
2 & $0.38^{+1}_{-1}$ & $0.38^{+1}_{-1}$ 
  & $0.38^{+1}_{-1}$ & $0.39^{+1}_{-1}$ \\
3 & $0.42^{+2}_{-2}$ & $0.41^{+2}_{-2}$ 
  & $0.41^{+2}_{-2}$ & $0.41^{+2}_{-2}$ \\
4 & $0.50^{+8}_{-9}$ & $0.48^{+8}_{-9}$ 
  & $0.45^{+9}_{-10}$ & $0.43^{+10}_{-10}$ \\
\hline \hline
\multicolumn{5}{c}{\bf Tadpole Improved} \\
\hline
$\Delta t$ & 0.154 & 0.155 & 0.156 
           & \multicolumn{1}{c}{$\kappa_c$} \\
\hline
2 & $0.49^{+1}_{-1}$ & $0.50^{+1}_{-1}$ 
  & $0.50^{+1}_{-1}$ & $0.50^{+1}_{-1}$ \\
3 & $0.55 ^{+3}_{-2}$ & $0.54 ^{+3}_{-2}$ 
  & $0.54 ^{+3}_{-2}$ & $0.53 ^{+3}_{-3}$ \\
4 & $0.65 ^{+10}_{-12}$ & $0.63 ^{+11}_{-12}$ 
  & $0.59 ^{+12}_{-13}$ & $0.57 ^{+13}_{-13}$ \\
\hline \hline \\
\multicolumn{5}{c}{\bf Renormalized Isgur-Wise Slope} \\ \hline \hline
\multicolumn{5}{c}{\bf Tadpole Improved} \\
\hline
$\Delta t$ & 0.154 & 0.155 & 0.156 
           & \multicolumn{1}{c}{$\kappa_c$} \\
\hline
2 & $0.57^{+1}_{-1}$ & $0.57^{+1}_{-1}$ 
  & $0.58^{+1}_{-1}$ & $0.58^{+1}_{-1}$ \\
3 & $0.62 ^{+3}_{-2}$ & $0.62 ^{+3}_{-2}$ 
  & $0.61 ^{+3}_{-2}$ & $0.61 ^{+3}_{-3}$ \\
4 & $0.72 ^{+12}_{-11}$ & $0.71 ^{+11}_{-11}$ 
  & $0.67 ^{+13}_{-11}$ & $0.64 ^{+13}_{-13}$ \\
\hline \hline
\end{tabular}
\caption{The negative of the slope at the normalization point,
	$\xi'(1)$, from both the unrenormalized and the renormalized
	ratio of three-point functions.  This ratio gives the
	(un)-renormalized Isgur-Wise function $\xi(v \!\cdot\!  v')$
	at asymptotically-large times $\Delta t$.}
\label{t:slope}
\end{center}
\vspace{-.25cm}
\end{table}

After renormalization of our tadpole-improved results, we obtain
$\xi^{\prime}_{\rm ren}(1) = -0.64(13)$ for the slope of the
Isgur-Wise function in continuum HQET in the $\overline{\rm MS}$
scheme at a scale of $4.0\,$GeV.  Without renormalization, the slope
is $\xi^{\prime\ {\rm tad}}_{\rm unren}(1) = -0.56(13)$.  Without
tadpole-improvement, the slope is $\xi^{\prime\ {\rm nt}}_{\rm
unren}(1) = -0.43(10)$.

We found that the tadpole-improved action (and therefore the
tadpole-improved data) cannot be obtained from the
non-tadpole-improved action (or data) by a simple rescaling of any
parameter.  However, the form of the evolution equation allows the
{\it construction\/} of the tadpole-improved Monte-Carlo data from
the non-tad\-pole-im\-prov\-ed data as described in
\S\ref{s:tadpole}.  After tadpole improvement, non-perturbative
corrections to the velocity are negligible.  Furthermore, tadpole
improvement greatly reduces the perturbative corrections to the slope
of the Isgur-Wise function.

\end{document}